\documentclass{moriond}
\usepackage{amssymb}
\def\be{\begin{equation}}
\def\ee{\end{equation}}
\def\bea{\begin{eqnarray}}
\def\eea{\end{eqnarray}}

\newcommand{\invfb}   {\mbox{$\rm fb^{-1}$}\ }
\newcommand{\mDM}   {\mbox{$m_{\rm DM}$}}
\newcommand{\mMED}   {\mbox{$m_{\rm MED}$}}
\newcommand{\gq}   {\mbox{$g_{\rm q}$}}
\newcommand{\gl}   {\mbox{$g_{\rm l}$}}
\newcommand{\gDM}   {\mbox{$g_{\rm  DM}$}}
\newcommand{\Zp}   {\mbox{$Z{\rm '}$}}

\newcommand{\Photo}{}

\begin{document} %
\vspace*{4cm} 
\title{SEARCHES FOR DARK MATTER PARTICLES AT THE LHC
}
\author{ MARTA FELCINI
\footnote{On behalf of the ATLAS and CMS Collaborations}
\footnote{marta.felcini@cern.ch}
}
\address{University College Dublin, School of Physics}

\maketitle\abstracts{
The searches for new particles that could be constituents of the dark matter in the universe are an essential part of the physics program of the experiments at the Large Hadron Collider. An overview of recent dark matter candidate searches is presented with a focus on new results obtained by the ATLAS and CMS experiments from the analysis of the proton-proton collision data at 13 TeV center-of-mass energy collected in the first part of Run~2.
}
\section{Introduction}
After the discovery of the Higgs boson at the CERN Large Hadron Collider (LHC), primary goals of the LHC physics program are the study with high precision of the Higgs boson interactions 
as well as the search for new particles and interactions that may explain some of the fundamental physics questions left unanswered so far.   One of the biggest questions is posed by the existence of the dark matter in the universe. While the evidence for dark matter (DM) is compelling, its nature is unknown. From the cosmological and astrophysical observational evidences, a picture has emerged that  supports the hypothesis of non baryonic DM.  
Well motivated candidates for non baryonic DM constituents are new elementary particles that are electrically neutral, stable and weakly interacting with  Standard Model (SM) particles. 
In this context,  three main experimental approaches are employed to probe the DM particle hypothesis and, in case of detection, to study its properties: (a) indirect detection (ID) experiments, with satellites and ground-based telescopes, searching for signals of DM annihilation in space, (b) direct detection (DD) experiments, searching for signal of nuclear recoil from  the scattering of DM particles in underground detectors, and (c) detection of DM particles production at colliders.

Recent results and perspectives for ID searches~\cite{Mijakowski,Behlmann,Boyarsky,Manconi}
and DD searches~\cite{Manfredini,Lehnert,Davini,Settimo,Bolognino,Gentile} have been presented at this Conference,  showing that  the present and upcoming experiments, 
with their unprecedented sensitivity,  
will detect cosmic DM particles or will set more stringent constraints on the current particle DM models.
The LHC searches are complementary to the ID and DD approaches in that they aim to 
probe SM-DM particle interactions at the present terrestrial energy frontier 
and possibly to measure the properties of the DM particles once detected. If a suitable DM particle candidate would be detected at the LHC,  DD and/or ID experiment should detect a signal from such particle interactions, to prove that it is a dominant component of the cosmic DM.
Conversely, if DD or ID experiment would detect a DM signal, new particle signals consistent with the observed ID and/or  DD signal could be detected and studied by the LHC experiments. 
In all cases, the combined results of the three approaches will have a deep impact on our understanding of the DM problem.

The LHC has started its second phase of operations (Run 2) in 2015 after two years of maintenance and upgrading. The center-of-mass energy in Run 2 is 13 TeV, a significant increase over the initial three-year LHC run (Run 1), which began with a center-of-mass energy of 7 TeV, rising to 8 TeV. 
The results presented here are based on the data collected by the ATLAS~\cite{Aad:2008zzm,ATLASALLPUBLIC}
and CMS~\cite{Chatrchyan:2008aa,CMSALLPUBLIC}
experiments in proton-proton (pp) collisions at 13 TeV. The integrated luminosity for this sample, mostly collected in 2016, amounts to about 36 \invfb per experiment.  
In 2017 an integrated luminosity of 50 \invfb has been delivered by the LHC to each of the two experiments and a similar  amount is expected to be delivered by the end of 2018, targeting a total Run 2 integrated luminosity of about 150~\invfb delivered to the experiments.
The energy and luminosity increase with respect to Run 1 allows to extend the experimental sensitivity 
to the production of new particles and rare new processes with cross-sections around and below one fb, as expected for many models of BSM physics.  

The sensitivity of the experiments to rare processes is illustrated by Fig.~1~(left)~\cite{ATLAS:2018lkl} showing a summary of several measurements of SM processes with cross-sections reaching down to values around and below one fb.  As shown in Fig.~1~(left) from ATLAS~\cite{ATLAS:2018lkl} (similar summary plots are also available 
from CMS~\cite{CMS:2018sum}), the LHC experiments have studied SM processes with cross-sections spanning over more than fifteen orders of magnitude in a large variety of final states and topologies. 
The SM processes are the `standard candles' of experimental particle physics.  Their  precise knowledge allows to search effectively either, in a model independent way,  for deviations from the SM expectations, or, based on specific models, for new signals of physics beyond the SM, for which the SM background can be precisely evaluated.  For all processes studied so far, the results are compatible with the SM predictions. 
A major research focus of the LHC experimental program is the study of the 125 GeV Higgs boson.
Fig.~1~(right)~\cite{CMS:2018lkl} gives an overview of the 125 GeV  Higgs production and decay measurements. The precise measurements of the Higgs boson  properties allow to test the SM predictions as well as to explore new physics avenues that could be opened up by the Higgs special nature (first spin-zero elementary particle to be observed), including the production of DM particle candidates through  invisible Higgs decays. 
\begin{figure}[tb]
\centering
\begin{minipage}{1.0\textwidth}
\includegraphics[width=1.0\textwidth]{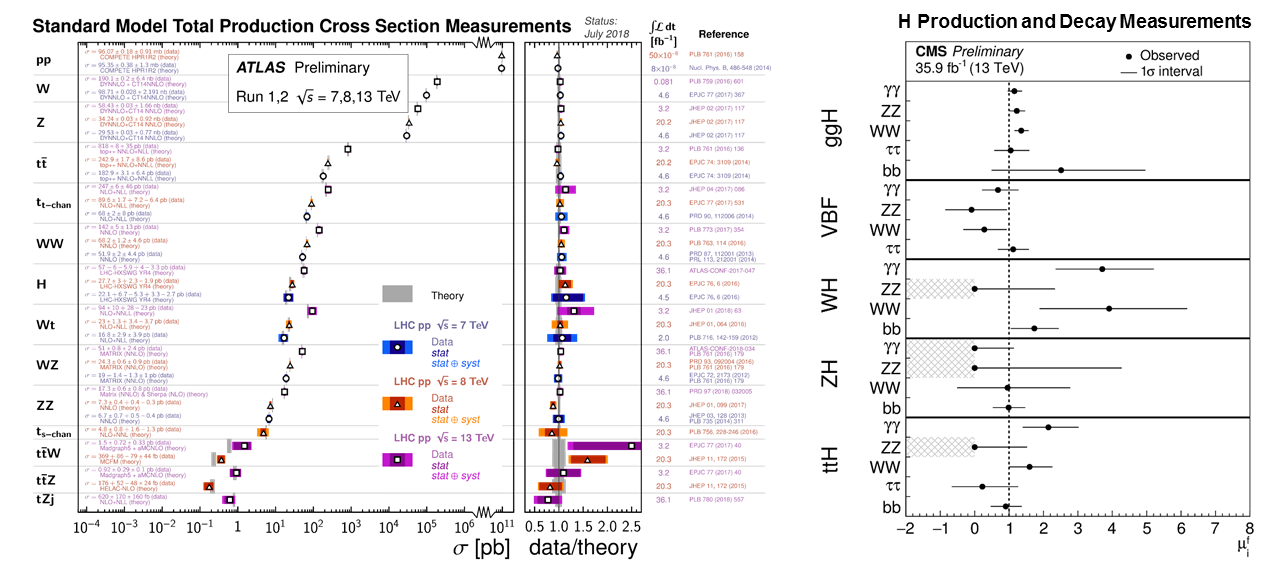}
\end{minipage}
\caption{Overview of (left) SM cross-section 
measurements~\protect\cite{ATLAS:2018lkl} compared to their SM expectation and (right) measurements~\protect\cite{CMS:2018lkl} of Higgs cross-sections and decay branching ratios normalized to their SM expectation.
}
\end{figure}

Several models have been proposed for the production in high energy collisions of new particles that would satisfy the cosmological and astrophysical constraints for a suitable DM particle candidate. 
Models of DM production at colliders range from a simple contact-interaction parametrization, 
with very reduced number of parameters, but of somewhat limited applicability, 
to minimal extensions of the SM including new boson fields mediating SM-DM interactions, to so-called ultraviolet (UV) complete models, such as Supersymmetry (SUSY).
Final states resulting from  the proposed DM production processes at colliders  
feature the presence of missing transverse momentum (also called missing transverse energy or MET), due to the DM particles interacting sufficiently weakly as to be invisible in the detector.  To be detectable, the DM production event must be accompanied by at least one visible high transverse momentum object (jet, lepton, photon, etc.), to trigger the acquisition of  the full event data. 
\begin{figure}[tb]
\centerline{\includegraphics[width=1.0\linewidth]{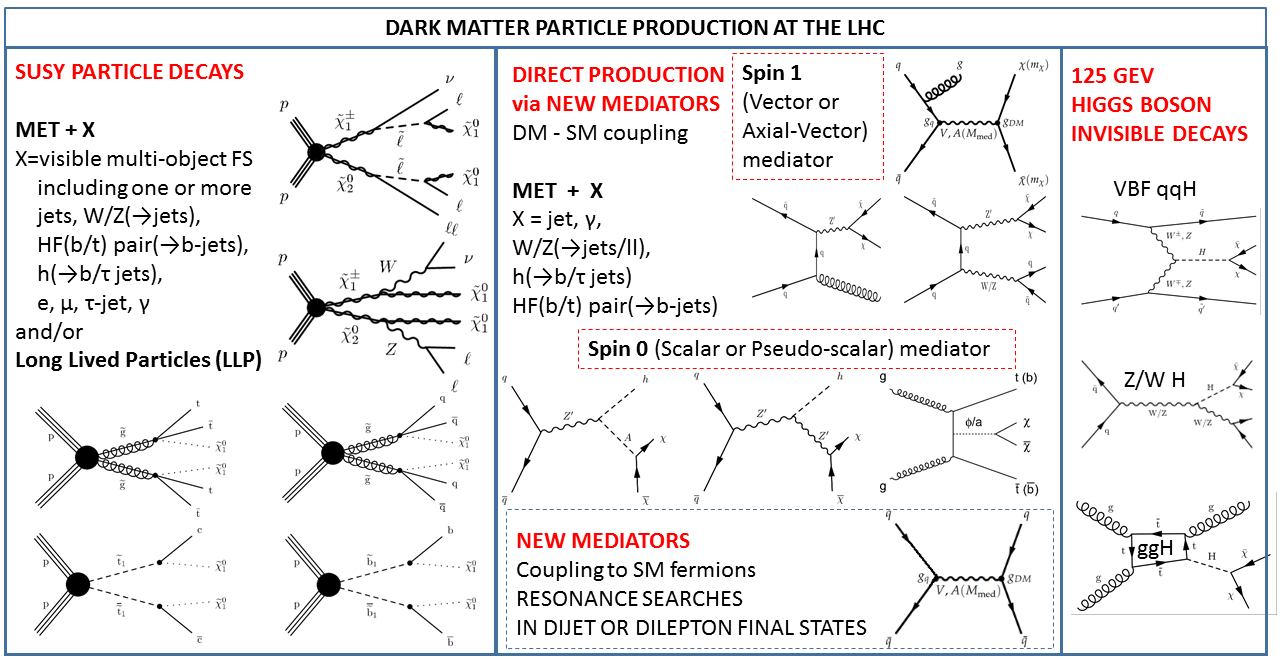}}
\caption{Examples of leading processes for DM particle production and searches at the LHC  for three main classes of models, shown from the left to the right panel in order of decreasing number of model parameters. 
}
\end{figure}

We classify these models in three main classes depending on the complexity of the phenomenology and the number of parameters involved in the interpretation of the experimental results. We discuss DM production in the decays of heavier SUSY particles, direct DM production through new mediators, and DM production in the decays of the 125 GeV Higgs boson.  Examples of DM production processes in the context of the three theoretical approaches are shown in Fig.~2. 
SUSY models (left), being the more complete from the theoretical view point, 
imply a large number of parameters~\footnote{As an example the Minimal Super Symmetric Model (MSSM) includes O(100) free parameters, while reduced versions, like the phenomenological MSSM, or  pMSSM, may imply $\sim$10 to $\sim$20 free parameters}.  
In simplified models of direct DM production through new mediators, 
the number of free parameters is limited to the DM and  mediator particle masses, spin/parity and couplings. 
The DM production through decays of the 125 GeV Higgs boson is the least parameter dependent of the three approaches as only three free parameters, the Higgs-DM coupling and the DM mass and spin,  enter in the description of the process. Since the DM particles are expected to be produced in pair, the invisible Higgs decay search is sensitive to DM particle masses up to half the Higgs boson mass.   

In the following, 
highlights of SUSY DM candidate searches are given in Section~2 while in Section~3  DM candidate searches in the framework of simplified model are discussed. In Section~4 results from the search for DM particles in Higgs decays are reviewed.  Finally Section 5 summarizes the conclusions.

\section{DM production in SUSY decays }
\begin{figure}[tb]
\begin{minipage}{1.0\linewidth}
\centerline{\includegraphics[width=1.0\linewidth]{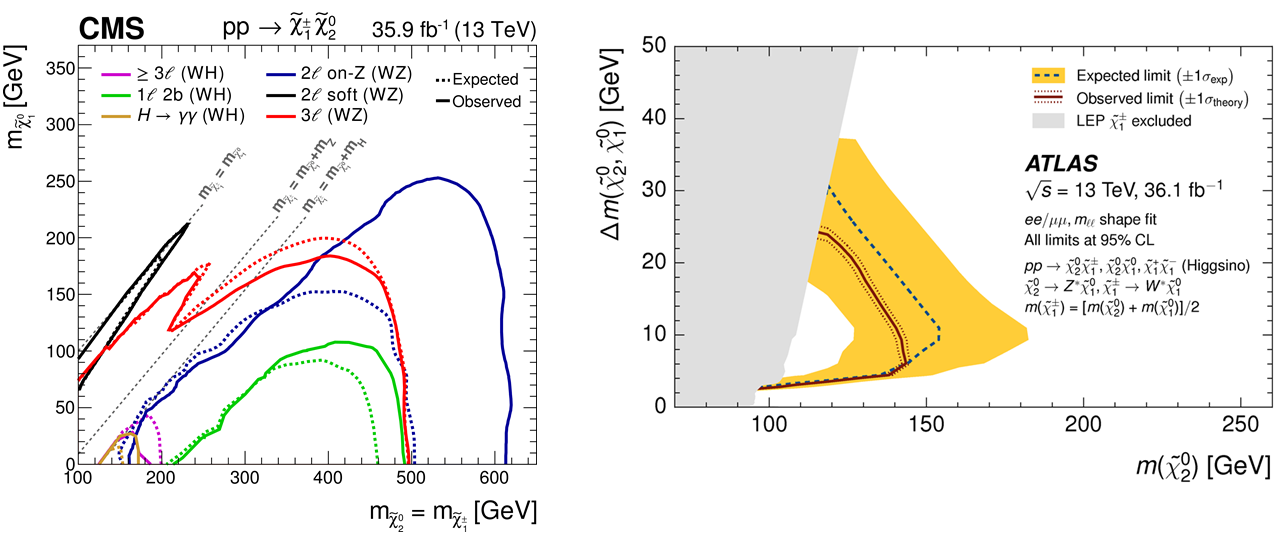}}
\end{minipage}
\caption{SUSY searches: results for (left) several chargino and neutralino 
searches~\protect\cite{Sirunyan:2018ubx}
and (right) in scenarios with compressed mass spectra~\protect\cite{Aaboud:2017leg}. 
 }
\end{figure}
SUSY models, as well as other UV complete models designed to solve the gauge hierarchy problem, may provide well motivated DM candidates. In particular,  SUSY models with  conserved  R-parity (RP)
predict the existence of a particle (typically the lightest SUSY particle or LSP)  that is stable, electrically neutral and feebly interacting with SM particles. Within these models, LSP's  are expected to be 
relics from the Big Bang~\cite{Ellis:2010kf} and thus could provide most of the observed DM in the universe. 
Under RP conservation, SUSY particles are produced in pair in the collision and, decaying promptly into lighter SUSY states plus SM particles,  result into final states containing several SM particles accompanied by a minimum of two invisible LSP's. The experimental signatures include MET plus multiple jets and/or leptons.  In models where the NLSP is long lived (because of suppressed decay to the LSP) the NLSP may leave the detector undetected, behaving as the LSP,  or leave a detectable signature. These cases are covered by long lived particle (LLP) searches. 

The findings of Run 1, with the discovery of one SM-like Higgs boson of 125 GeV mass, and the lack of observation of strongly interacting SUSY particles with masses above one TeV (in the context of the constrained MSSM), have motivated searches for relatively lighter particles produced with substantially smaller cross-sections (justifying why they have not been observed yet). For SUSY models to provide a "natural" solution in stabilizing the Higgs mass and solving the hierarchy problem 
(for a review and original references, see {\it e.g.} Ref.~\cite{Feng:2013pwa}), 
there should exist relatively light chargino/neutralino states (particularly the Higgsino) and  third generation (top, bottom) squarks.
Expected final states may be particularly complex and challenging for detection, as it is the case for compressed mass spectra, when the mass difference between the NLSP and the LSP is small, thus producing events with little visible energy, requiring dedicated searches.  When decays of the NLSP into the LSP are suppressed, e.g.  because of very compressed mass spectra, production of LLPs is an additional possibility, thus requiring specific detection techniques. Similar topologies are expected in models with the LSP being the gravitino and the NLSP being a LLP. 

The LHC experiments conduct a large variety of SUSY particle searches\cite{atlassusypublic,cmssusypublic}.
Discovery (at the 5 sigma level) of any of the predicted signals would result into the measurement of a cross-section times branching ratio for SUSY processes consistent with the observation. After the discovery, significantly more luminosity would be needed to establish the properties of the particles involved in the observed process, in particular a precise measurement of  the LSP mass. In the absence of a signal, results are often provided  in terms of cross-section upper limits, for a given production and decay process, as a function of the  mass of the heaviest SUSY particle involved in the sought process and  the LSP mass. The detection efficiencies are evaluated in the framework of simplified models where the heaviest SUSY particle decays into the LSP and associated visible final states in one or two decay stages. The experiments provide efficiency maps and other detailed information so that the results can be recast into models predicting multi-staged decay chains.  

We show here results for chargino/neutralino searches and for t/b/q squarks and long lived gluino searches. 
Fig.~3 shows the results of the searches for chargino pair  production and for associate production of chargino 
and next-to-lightest neutralino.  A comparison of constraints~\cite{Sirunyan:2018ubx}
obtained from a number of different CMS searches in multi-leptons final states is shown in Fig.~3~(left). Similar constraints are reported by 
ATLAS~\cite{atlassusysummary201803} 
with most recent results from Ref.~\protect\cite{Aaboud:2018jiw}.
The region of the LSP-NLSP mass plane close to the diagonal, 
where the mass difference between the NLSP and LSP is less than 50 GeV, 
in the so-called compressed chargino/neutralino mass spectra scenarios, 
is covered by searches for chargino/neutralino production in final states 
with soft leptons~\cite{Aaboud:2017leg,Sirunyan:2018iwl}.  Fig.~3~(right)~\cite{Aaboud:2017leg} shows that for extreme cases where the mass difference is below 2 GeV, these  results do not improve yet  the LEP constraints.
Many searches  for production and decays of t/b/q squarks or gluinos are also conducted by the LHC experiments. 
The possibility of long lived gluinos, decaying far from the interaction region,  requires dedicated searches. 
\begin{figure}[tb]
\begin{minipage}{1.0\linewidth}
\centerline{\includegraphics[width=1.0\linewidth]{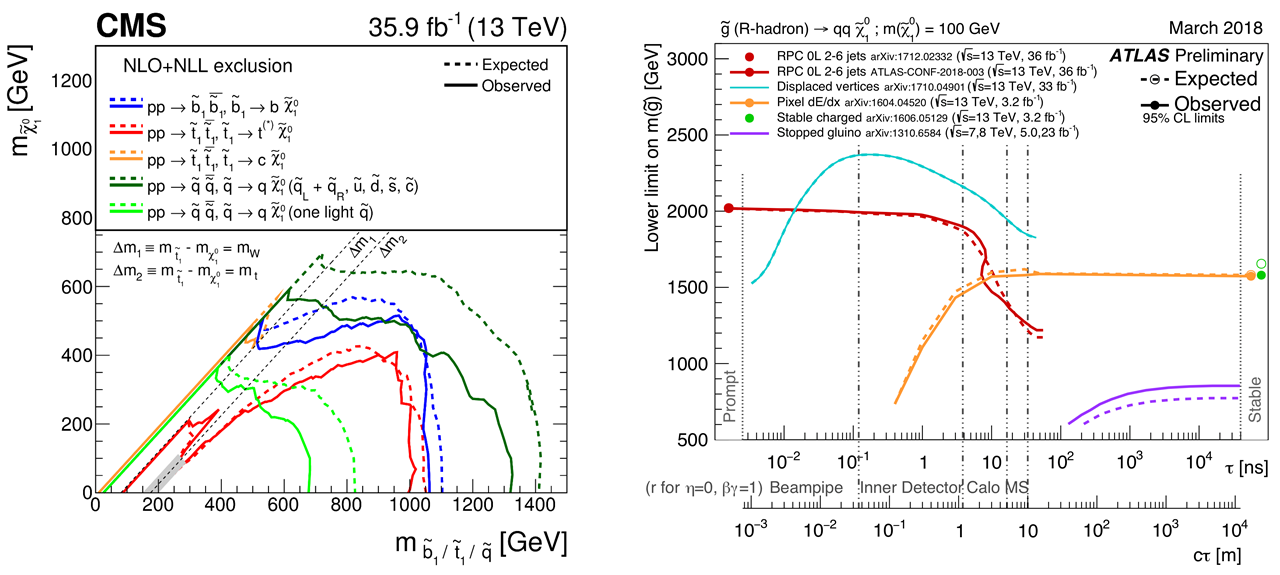}}
\end{minipage}
\caption{SUSY searches: results for (left) top, bottom, light  flavor squark searches~\protect\cite{Sirunyan:2018vjp} and (right) 
long lived gluino searches~\protect\cite{atlassusysummary201803} including recent results from Ref.~\protect\cite{Aaboud:2017vwy,ATLAS:2018yey,Aaboud:2017iio}.
}
\end{figure}
Fig.~4~(left) shows the results from CMS searches for t/b/q squarks~\cite{Sirunyan:2018vjp}.
ATLAS recent results on t/b/c squark searches are documented in 
Ref.~\cite{Aaboud:2017vwy,Aaboud:2017aeu,Aaboud:2018zjf}.
Fig.~4~(right) \cite{atlassusysummary201803}, including recent results from Ref.~\protect\cite{Aaboud:2017vwy,ATLAS:2018yey,Aaboud:2017iio}, shows mass constraints for a gluino with a long lifetime, as the gluino decay proceeds through highly  virtual squarks, the latters being much heavier than the gluino itself.
Recent results on long lived gluinos searches are also reported by CMS~\cite{Sirunyan:2018vjp,Sirunyan:2017sbs}. 

SUSY scenarios  with compressed mass spectra, but also other SUSY models (Split, GMSB, RPV SUSY etc), as well as many other models of BSM physics,  predict the existence of new LLP's. These particles are produced in the collision but do not decay promptly at the interaction region. They may decay inside the detector, with their decay products being measured in one or more of the sub-detectors, or they may decay outside of the detector (in this case they are labeled as stable, charged or neutral, particles) leaving or not a detectable trace.  Expected signatures are depending on the lifetime and other properties of these particles.  They can be displaced vertexes or disappearing tracks in the tracking devices, close to the interaction region, or displaced jets or leptons or photons, leaving signals also in the calorimeters, and in the muon chambers. Heavy stable charged particles cross the full detector with tracks in the tracking devices and energy loss in the calorimeters. 
Efforts to search for new LLP signals in the 13 TeV data are focused on developing dedicated techniques, from triggering, to event reconstruction to background estimation methods~\cite{Sirunyan:2018vjp,Sirunyan:2017sbs,Aaboud:2017mpt}. 

Exploiting the large amount of results from SUSY searches to constrain the allowed parameter space  is 
of particular importance for future research directions. Global statistical combinations including large sets of different search results have been performed. For example, it is shown~\cite{Bagnaschi:2017tru} 
that in the pMSSM11, with 11 free parameters, SUSY DM-nucleon  scattering cross-section values in the range to be explored by present and forthcoming DD experiments, for DM masses between $\sim 100$ GeV and 1 TeV, are still allowed. 
\section{Direct DM particle production via new DM-SM mediators}
\begin{figure}[tb]
\begin{minipage}{1.0\linewidth}
\centerline{\includegraphics[width=1.0\linewidth]{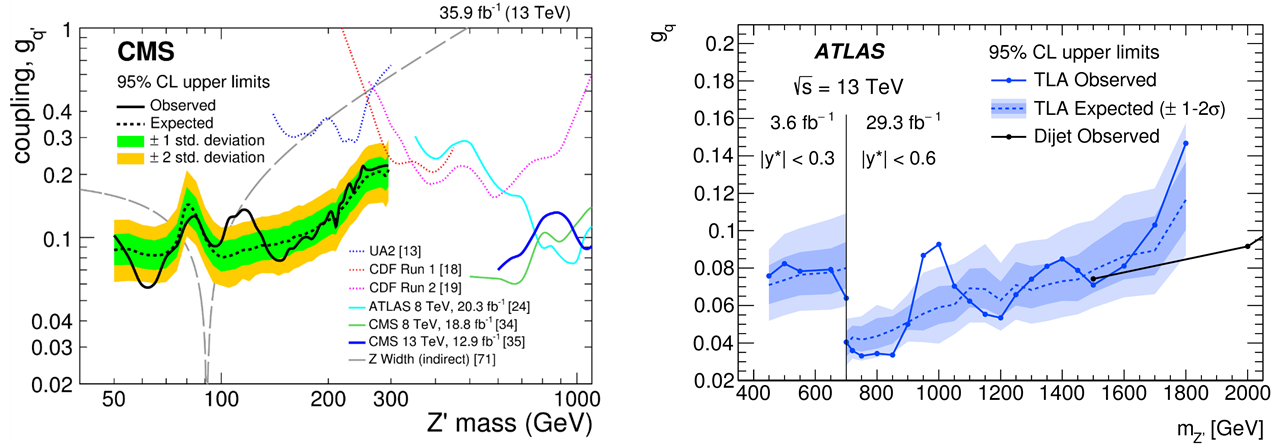}}
\end{minipage}
\caption{Low-mass dijet resonance search results from (left)~CMS~\protect\cite{Sirunyan:2017nvi} in the range 50-300 GeV using 
jet substructure techniques and 
(right)~ATLAS~\protect\cite{Aaboud:2018fzt} in the range 450-1800 GeV using trigger-level jets.}
\end{figure}
Another approach in the search for DM particles at the LHC is proposed by simplified models~\cite{Abdallah:2015ter,Boveia:2016mrp,Albert:2017onk}  of DM particle production at colliders, based on the idea that,  
in a generic theory of physics beyond the SM, most new particles may be too heavy to be produced at the present collision energy or even to play a role via virtual effects. Rather, the only kinematically accessible new states are a DM particle with a sizable interaction with the SM fermions and a new boson that mediates the SM-DM interaction.  
The mediator can be either a spin 1 (vector or axial-vector) or a spin 0 (scalar or pseudo-scalar) boson while the DM particle is assumed to be a Dirac fermion. Thus the DM-SM interaction is depending on few physics parameters:  the DM mass \mDM, the mediator mass \mMED\  and spin/parity, the mediator couplings \gq\ and \gl\  to SM quarks and leptons, respectively, and the mediator coupling  \gDM\  to DM particles. 
These models
 are used to design DM searches at the LHC, interpret their results and provide a common framework for comparison to non collider search results. 
 
 Relatively simple final states  (see Fig.~2, central panel) are expected, with a pair of invisible DM particles, leading to large MET,  produced in association with visible SM particles 'X'. Recent 'MET+X' searches include: 
MET+photon~\cite{Aaboud:2017dor,CMS:2018zwz}, 
MET+jet/V(jj)~\cite{Aaboud:2017phn,Aaboud:2018xdl,Sirunyan:2017jix}, 
MET+Z(ll)~\cite{Aaboud:2017bja,Sirunyan:2017qfc}, 
MET+H(bb)~\cite{Aaboud:2017yqz} or H($\gamma\gamma$ or $\tau\tau$)~\cite{Sirunyan:2018fpy},  MET+HF(b,t) quark pair~\cite{Aaboud:2017rzf,Sirunyan:2018dub}.
An additional prediction of simplified DM models is that mediator production and decay into SM fermions should lead to the detection of new resonances. While the search for high mass resonances (above one TeV)  has been a flagship search at the LHC, the simplified DM models have drawn attention to the fact that relatively light resonances (below or about one TeV) should also be searched for, as they may have gone undetected so far because of the relatively weak couplings to SM particles. The search for new low mass resonances, in particular in dijet final states,  is particularly challenging at the LHC because of the large SM backgrounds. Search methods have been developed that utilize new techniques both at the trigger level as well as in the off-line analysis. 
Fig. 5 shows results from low mass resonance searches in dijet final states. They are presented in terms of upper limits on a  mediator (\Zp) coupling to quarks as a function of the mediator mass: (left) 
in the range $50-300$ GeV using jet substructure techniques~\cite{Sirunyan:2017nvi},  to identify jets consistent with a particle decaying into a quark pair,  and (right) 
in the range $0.45-1.8$ TeV using inclusive low transverse momentum jet triggered events~\cite{Aaboud:2018fzt}, recording only limited event information, thereby allowing high event rates and search efficiency with reduced storage needs.
Other recent searches for mediators using Run 2 (13 TeV) data have been performed to cover a resonance mass range from 70 GeV to 7 TeV, exploiting diverse techniques: (i) in the low mass range, up to about 1 TeV, using large-radius jets~\cite{Aaboud:2018zba}, with  jet substructure identification technique,  or using dijets reconstructed 
at the trigger level~\cite{Sirunyan:2018xlo}  from calorimeter information,  and (ii) in the high mass region, exploiting  the dijet invariant mass and angular distributions~\cite{Aaboud:2017yvp}, using dijets reconstructed off-line with a particle-flow algorithm~\cite{Sirunyan:2018xlo}, and jets containing b-hadrons~\cite{Aaboud:2018tqo}. Searches in dilepton final states~\cite{Aaboud:2017buh,Sirunyan:2018exx} cover the mass range from 70 GeV to 6 TeV.
\begin{figure}[tb]
\begin{minipage}{1.0\linewidth}
\centerline{\includegraphics[width=1.0\linewidth]{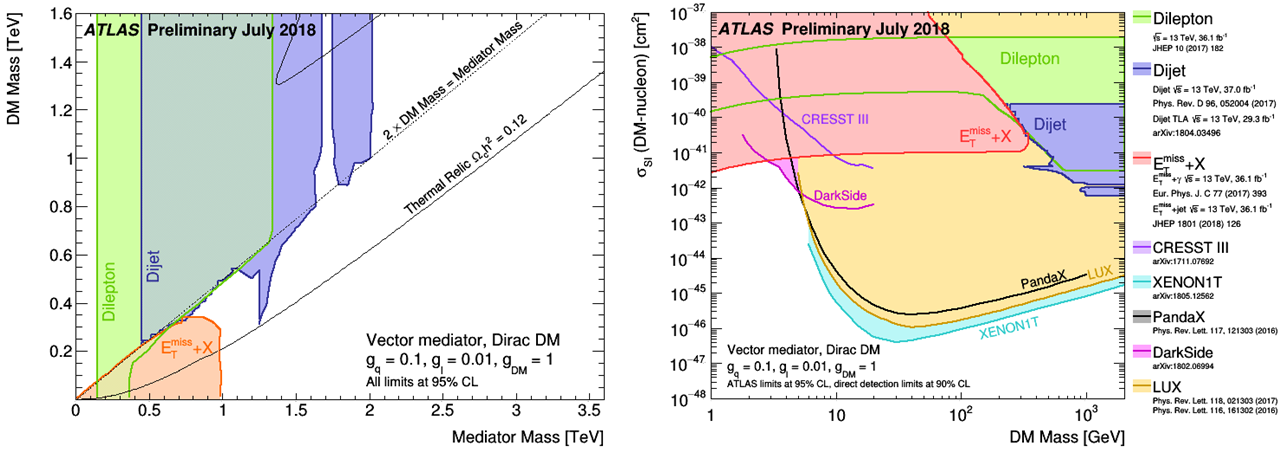}}
\end{minipage}
\caption{MET+X and dijet/dilepton search results interpreted~\protect\cite{atlasdmsummaryplots201807} in the context of simplified models: (left) excluded regions in the  $\mDM-\mMED$  plane and (left) upper limits on the DM-nucleon scattering cross-section as a function of \mDM, compared to results from DD experiments.
 }
\label{fig:ATLASDMINTERP}
\end{figure}
%

If no signal is detected  the LHC experiments set upper limits on the cross-section times branching ratio of the process sought for,  as a function of \mDM\ and \mMED\ for fixed values of the mediator spin/parity and \gDM, \gq\ and \gl\ couplings. An example is shown in Fig. 6 (left)~\cite{atlasdmsummaryplots201807}.  The region for $\mMED\geq 2\mDM$, where DM pair production takes place,  is covered by MET+X searches. The regions excluded by these searches in the $\mDM - \mMED$ plane have an approximately triangular shape, extending to a maximum excluded \mMED\ value (about 1 TeV for the coupling choices of Fig.~6) and, for an excluded  \mMED\ value, covering \mDM\  masses up to about \mMED/2. For smaller values of the couplings, the $\mDM-\mMED$ excluded region would be reduced. 
%
In the case of the resonance searches,
the excluded regions  in the $\mDM - \mMED$ plane are typically vertical stripes excluding a range of \mMED\  values, almost inedependently of \mDM, except in cases when the \gq\ or \gl\ values are so small that, 
in the region $\mMED\geq 2\mDM$,  the DM pair production is dominant and the mediator decay to SM fermions is largely suppressed. This is the case in Fig.~6~(left) where 
\gq\  and \gl\ 
are taken to be 0.1 and 0.01, respectively. The dijet searches cover the vertical region  for $\mMED< 2\mDM$,  while the dilepton searches have some coverage also in the $\mMED\geq 2\mDM$. Results for different mediator spin/parity and coupling choices are publicly available from ATLAS~\cite{atlasdmsummaryplots201807}
and CMS~\cite{cmsdmsummaryplots201807}.
%
%
In the framework of these simplified models, 
the LHC $\mDM-\mMED$ excluded regions can be recast into 
upper limits on the DM-nucleon scattering cross-section as a function of \mDM. 
In the case of DD experiments the DM-nucleon scattering cross-section upper limit is independent of \mMED, as the momentum transfer involved in the DM scattering off the target nucleus, is much smaller than the hypothetical mediator mass and \mMED-related effects are negligible.  The DM-nucleon scattering cross-section limits set by the LHC searches depend also on the assumed \mMED\ value insofar that, together with the coupling values,  it determines whether the DM pair production, covered by MET+X searches,  is the dominant process. 
For the above-mentioned choice of mediator couplings, the LHC limits, compared to DD results, are shown in Fig. 6~(right)~\cite{atlasdmsummaryplots201807}. Regions excluded by the MET+X searches and by the dijet and dilepton searches are shown separately reflecting directly the excluded $\mDM-\mMED$ regions in Fig. 6~(left). It is important to emphasize that these limits are only valid in the framework of the considered models and for the specific parameter choice. On the other hand,
DM searches in the framework of simplified  models  are inclusive enough to cover a broad range of topologies, 
which may arise in more complex scenarios, so that these searches results can be interpreted to constrain other models. 
Extending the searches to other possible processes, e.g. involving DM in association with Higgs-like or long lived particles, may provide a more complete framework to interpret and relate collider to non collider DM search results.

\section{DM production in the decay of the 125 GeV Higgs boson}
In the SM, the Higgs boson decays invisibly only through the ZZ decay to four neutrinos 
with a branching ratio of about 0.1\%.  The Higgs invisible decay rate may be largely enhanced in the context of BSM scenarios, in particular if the Higgs boson decays to DM particles. Indirect constraints on the Higgs invisible decay branching ratio can be inferred from the measurements of the visible decay channels: an upper limit of 34\% have been  obtained from a combination~\cite{Khachatryan:2016vau} of Higgs visible decay measurements using Run 1 (7-8 TeV) data.
More recently, direct searches for invisible Higgs decays using Run 2 (13 TeV) data have been performed 
targeting
the vector boson fusion channel~\cite{CMS:2018awd}, in which the Higgs boson is produced in association with jets (VBF via qq$\rightarrow$qqH),
 the associated production of a Higgs boson with 
 Z/W (Z$\rightarrow$ll, Z/W$\rightarrow$jj) modes~\cite{Aaboud:2017bja,Sirunyan:2017qfc,Sirunyan:2017jix,Aaboud:2018xdl}
and the ggH production channel, where a high pT Higgs boson is produced in association 
with initial state radiation jets~\cite{Sirunyan:2017jix}.
Leading Feynman diagrams for the qqH, VH, and ggH processes are shown in Fig.~2~(left panel).
\begin{figure}[tb]
\begin{minipage}{1.0\linewidth}
\centerline{\includegraphics[width=1.0\linewidth]{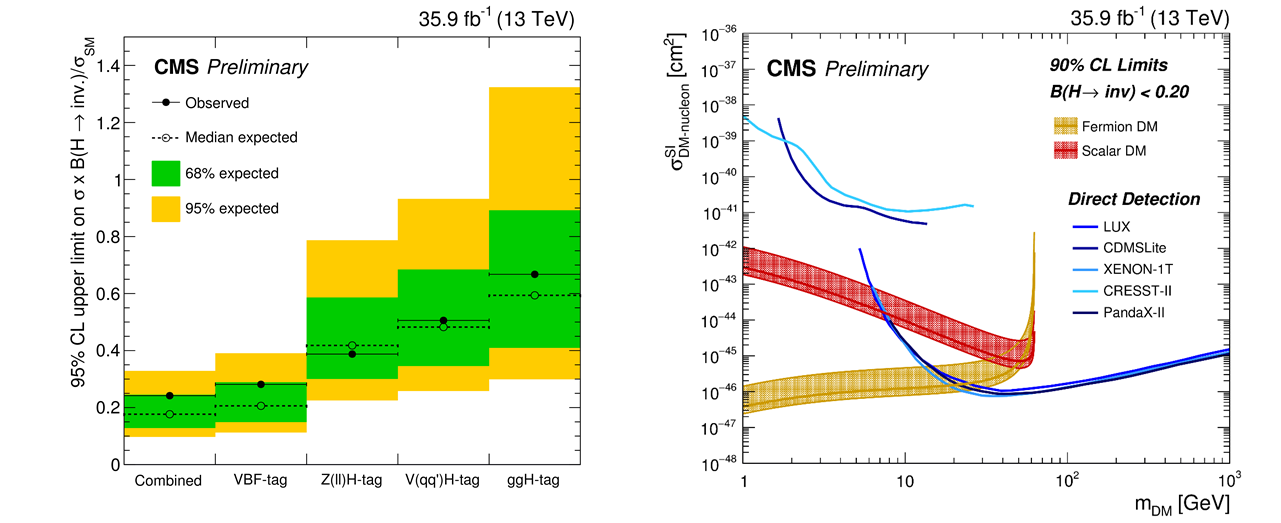}}
\end{minipage}
\caption{Higgs invisible decay searches~\protect\cite{CMS:2018awd}: (left) upper limits on the relative invisible decay rate for individual searches and the combination and (right) upper limits on  the DM-nucleon scattering cross-section assuming a scalar or fermion DM candidate, compared to limits from DD experiments. 
}
\label{fig:CMSINVHIGGS}
\end{figure}
No significant deviations from the SM predictions are observed in any of these searches. 

These results are interpreted in terms of  upper limits on the product of the Higgs
production cross-section and branching ratio to invisible particles, $\sigma$B(H$\rightarrow$inv),
relative to the SM Higgs production cross-section $\sigma_{\rm SM}$.
Observed and expected upper limits on $\sigma$B(H$\rightarrow$inv)/$\sigma_{\rm SM}$ 
at 95\% CL are presented in Fig.~7~(left)~\cite{CMS:2018awd} for each search channel, with the VBF being  the most sensitive search. 
The statistical combination including these search results  
yields an observed (expected) upper limit on 
B(H$\rightarrow$inv)
of 0.24 (0.18) at 95\% CL, assuming
SM Higgs production cross-section. 
The observed 90\% CL upper limit  on  B(H$\rightarrow$inv)
of 0.2 is interpreted in the context of a Higgs-portal model of DM interactions to set a 90\% CL upper limit on the spin-independent DM-nucleon interaction cross-section as a function of the DM mass, shown in Fig.~7~(right)~\cite{CMS:2018awd}. In direct comparison with the corresponding upper limits from DD experiments, it  provides the strongest constraint on fermion (scalar) DM particles for DM masses smaller than 20 (7) GeV.
\section{Conclusions}
Many searches for DM particle production have been conducted by the LHC experiments and constraints derived in the framework of complete BSM models, such as SUSY, and of simplified DM production models.   
Upper limits on DM production cross-sections are derived as function of the DM particle mass. For simplified models, in the search for DM production mediators, depending on mediator couplings, mass constraints extend in the range up to few TeV. 
Attention has been raised on the fact that simplified models may set too optimistic limits for cases when complex multi-object final states (with {\it e.g.} multiple W/Z's or t/b quarks)  and more exotic signatures are expected. Experiments are investigating more complex possibilities developing new search methods and new tools. An additional mean to search for DM particles is through the invisible decay of the 125 GeV Higgs boson. This channel has access to small DM particle masses up to one half the Higgs mass. 

An integrated luminosity of about 150 \invfb is expected to be delivered by the LHC until the end of Run 2 in 2018, more than a factor of three larger than the sample for which results have been reported here. During Run 3, starting in 2021,  the LHC experiments are expected to collect 300 \invfb\ at 14 TeV. In the High Luminosity phase 
of the LHC (HL-LHC)~\cite{lhcscheduke201803}
an integrated  luminosity of at least 3000 \invfb, will be collected, representing a  factor of more than 20 in the statistics of potential new signals as compared to Run 2. Experiments have engaged in substantial detector and  trigger upgrades to maintain, and possibly improve,  the present detector performance in the harsh running conditions 
imposed by the high number ($\sim$200 in average) of multiple interactions (pile-up) per bunch-crossing expected 
at the HL-LHC.

Overall, there is still ample room for discoveries of DM particle candidates at the LHC. 
This is especially relevant for the observation of weakly ({\it i.e.} rarely) produced particles, 
in particular, for the lightest neutralino, for which the current mass limits in some cases do not improve on the LEP constraints yet. Extending  the search sensitivity to  the weakly produced particle mass range in the few hundreds GeV's and beyond is a strong motivation for the HL-LHC.
\section*{Acknowledgments}
I would like to thank the organizers of the Rencontres de Moriond Cosmology for inviting the LHC experiments to present their results to this view-broadening and inspiring conference, for the warm hospitality and the friendly atmosphere.
I am grateful to my  ATLAS and CMS colleagues for giving me the opportunity to present this talk on behalf of the Collaborations. 
%
\section*{References}

\newcommand*{\doi}[1]{\href{http://dx.doi.org/#1}{doi:#1}}
\newcommand*{\arx}[1]{\href{https://arxiv.org/abs/#1}{arXiv:#1}}
\newcommand*{\arxhepex}[1]{\href{https://arxiv.org/abs/#1}{arXiv:#1[hep-ex]}}

\end{document}